\documentclass[runningheads]{llncs}
\usepackage{graphicx}
\usepackage[T1]{fontenc}

\begin{document}

\title{Artificial Intelligence Technologies in Education:
Benefits, Challenges and Strategies of Implementation}
\titlerunning{Artificial Intelligence Technologies in Education: (...)}
\author{Mieczysław L. Owoc\inst{1}\orcidID{0000-0003-1578-6934} \and
\\ Agnieszka Sawicka\inst{1}\orcidID{0000-0002-8318-9931} \and
\\ Paweł Weichbroth\inst{2}\orcidID{0000-0002-1645-0941}
}

\authorrunning{M. L. Owoc et al.}
\institute{Wrocław University of Economics and Business, \\ Komandorska 118/120 street, 53-345 Wrocław, Poland \\
\email{\{mieczyslaw.owoc;agnieszka.sawicka\}@ue.wroc.pl}\\
\url{http://www.ue.wroc.pl/en/}
\and
Gdańsk University of Technology,\\ Faculty of Electronics, Telecommunications and Informatics,\\
Department of Software Engineering, \\ 11/12 Gabriela Narutowicza Street, \\80-233 Gdańsk, Poland\\
\email{pawel.weichbroth@pg.edu.pl}\\
\url{https://pg.edu.pl/en/home}}

\maketitle
\begin{abstract}
Since the education sector is associated with highly dynamic business environments which are controlled and maintained by information systems, recent technological advancements and the increasing pace of adopting artificial intelligence (AI) technologies constitute a need to identify and analyze the issues regarding their implementation in education sector. However, a study of the contemporary literature reveled that relatively little research has been undertaken in this area. To fill this void, we have identified the benefits and challenges of implementing artificial intelligence in the education sector, preceded by a short discussion on the concepts of AI and its evolution over time. Moreover, we have also reviewed modern AI technologies for learners and educators, currently available on the software market, evaluating their usefulness. Last but not least, we have developed a strategy implementation model, described by a five-stage, generic process, along with the corresponding configuration guide. To verify and validate their design, we separately developed three implementation strategies for three different higher education organizations. We believe that the obtained results will contribute to better understanding the specificities of AI systems, services and tools, and afterwards pave a smooth way in their implementation.

\end{abstract}
\section{Introduction}
Research into artificial intelligence (AI) has tended to increase for a number of reasons. The desire and excitement to create intelligent systems was the case in the past in the academic community, while now it seems inevitable that their development and inception have been addressed by the majority of organizations, representing almost all business sectors. It is not only the effect of the far-reaching technological progress such as microprocessors, data storage and global networking, but also the impact of changes in business strategies. While the debate on how AI will change business is at the top of the present-day agenda [1], education is already being challenged to reconceptualize existing teaching and learning methods by putting AI techniques and tools into service [2]–[4].

Indeed, Sameer Maskey, founder and CEO at Fusemachines, an AI Education and AI Talent Solution provider based in New York City (USA), in an interview with Forbes magazine published on 8th June 2020, said that [5]:
“It will be important for educators and policymakers to explore the intersection of education and artificial intelligence. The application of machines in learning environments is only one variable in a multifaceted equation. We have to consider barriers that prevent an even distribution in technological resources and how to overcome them. We must also ensure that teachers are prepared and empowered to leverage artificial intelligence. Assuming these elements are addressed, the possibilities of AI-powered learning are infinite”.

Inspired by his words, and apart from the financial costs and profits, the question arises what are the specific benefits and challenges of implementing artificial intelligence in education? Since little work has focused on this area, the first research goal is to identify and analyze its possible benefits and preceding challenges (see Sec. 2.2 and 2.3, respectively). In this case, a qualitative generic thematic analysis was undertaken due to its flexibility to collect descriptive data in the narrative form.

Our study is also driven by a concern for the implications of the human factor due to the increasing evidence of the apprehension raised against the proliferation of artificial intelligence since the variety of its applications seems to be immeasurable, ranging from computer games [6,7], decision-making [8]–[10], education [11]–[16], enterprise management [17]–[19], grid computing [20]–[23], knowledge management [24]–[27], learning systems [28]–[31], ontologies [32]–[34], smart cities [35]–[37], and software engineering [38]–[40], to name just a few. On the contrary, we attempt to argue that AI can amplify educational effectiveness, concerned with sharing, developing and disseminating knowledge, at the same time preserving human autonomy, agency and capabilities. Therefore, the second research goal is to analyze and evaluate the usefulness of existing AI technologies (see Sec. 2.4).

To explore these two goals, we collect, analyze and review a plethora of information sources. Their addresses were obtained from the Google web search engine and Google Scholar website by using combinations of keywords such as: “artificial intelligence”, “adoption”, “acceptance”, “advantages”, “disadvantages”, “implementation”, “deployment”, “education”, “benefits”, “challenges”, “risks”, and “technologies”.

What is much rarer, however, was to find applicable and relevant implementation strategies. Nevertheless, based on our previous results, which provide solid theoretical foundation, we also designed and created a generic strategy, able to be applied in a strategic plan regarding any of the AI systems, services or tools in all industries, and in organizations of a variety of sizes. In particular, our strategy addresses the “what” and “why” of the activities, embedded in a five-phase process model (see Sec. 3).

To verify and validate its design, we separately developed three implementation strategies (see Sec. 4), for three different non-public higher education institutions, namely the “Copernicus” Wroclaw Computer Science and Management University (WSZI), WSB University in Gdansk (WSB) and Jan Wyżykowski University (UJW). By adopting a qualitative study design, as an exploratory, descriptive approach, in the first step we collected all of the necessary data, using a specific thematic approach, while in the second step we analyzed the data in a fashion reflecting the aim of recognizing circumstances and challenges related to the subject matter.

It is worth noting here that nowadays, non-public universities play an essential role in higher education [41]–[43] and have to compete very hard in order to gather potential students [44]–[46]. Their position increasingly depends on the quality of the education and the managerial competencies of the university governance [47]–[49]. In both cases, applying intelligent technologies seems to be a must if one considers their competitiveness and development. Yet, the level of its implementation is still relatively low in comparison with the business sector.

On the other hand, there are a few cases documented, giving an idea in which areas AI methods have been implemented within higher education institutions (HEIs) [50]–[52]. In particular, intelligent technologies are gradually being implemented in non-public universities [53], usually being part of the strategy that sets up a framework of priorities [54]. However, to the best of our knowledge, very few studies have considered the benefits and challenges affecting the implementation of AI technologies within university emerging set-ups.

The rest of the paper is structured as follows. Section 2 discusses artificial intelligence (AI) in the education sector, introducing the basic concepts, benefits and challenges related to its implementation, as well as a discussion of modern AI systems and tools. Section 3 presents a strategy implementation model, conceptualized by a five-stage generic process along with the corresponding configuration guide. Section 4 describes and analyzes the case studies, illustrating how AI implementations, based on underlying decision processes, are conducted in practice. Finally, we conclude in Section 5.

\section{AI in Education}
By design, intelligent technology is a method which uses knowledge to achieve concrete purpose in efficiency. At present, there are the following intelligent technologies: multi-agent, machine learning, ontology, semantic and knowledge grid, autonomic computing, cognitive informatics and neural computing. The prompt advances in these fields have already brought substantial changes in education, opening up new opportunities and challenges to teach and learn anytime and anywhere by providing new methods and systems that aim to stimulate innovative teaching and ultimately improve learning outcomes.
\subsection{2.1 AI concepts}
\begin{figure}
\centering
\includegraphics[height=6.2cm]{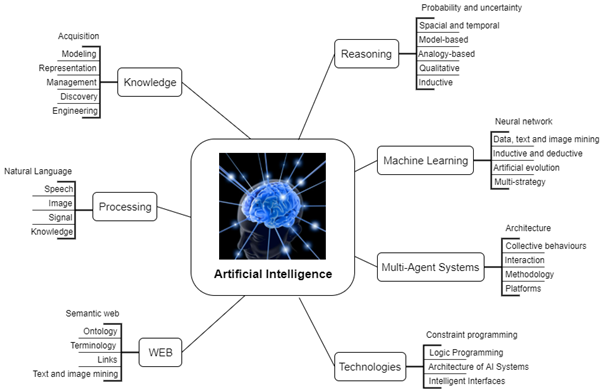}
\caption{Artificial Intelligence 3.0 [55].}
\label{fig:AI30}
\end{figure}

The continuous progress of modern information technologies is strictly connected with the presence of implemented artificial intelligence techniques. During the over 60 years of development of artificial intelligence, several intelligent approaches have appeared in almost all sectors of modern life. Therefore, one can talk about the new generation of AI, including the potential power of the current solutions and the variety of applied techniques. The crucial components of such an understanding of AI 3.0 are presented in Figure \ref{fig:AI30}.

Particular categories of AI can be combined in the final applications; some of them seem to be obligatory (knowledge, reasoning, processing) while others are employed for the specific solutions where knowledge should be permanently updated (machine learning) or requires cooperation between specialized agents (multi-agent systems). Either way, the current solutions are not quite satisfactory – the level of the obtained progress is less than human intelligence. The next imaginable stages are still before us – we are heading toward Artificial Super Intelligence passing in the meantime through Artificial General Intelligence (see Fig. 2).
\begin{figure}
\centering
\includegraphics[height=6.2cm]{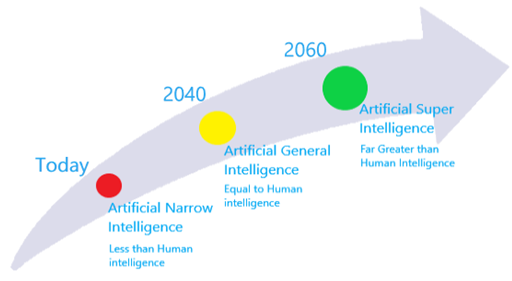}
\caption{Future evolution of Artificial Intelligence [56].}
\label{fig:EvolutionAI}
\end{figure}

To sum up, the landscape of Artificial Intelligence in terms of using the main categories is rather stable. Yet, learning-based techniques are playing an increasingly significant role. However, the specialty of the application areas can determine the shape of implementation of artificial intelligence methods. The educational sector and the specifics of non-public universities are real determinants in defining the implementation of intelligent technologies.

\subsection{AI technologies}
With the advent of AI in the mid-1950s to the present day, the proliferation of its methods and techniques has made it possible to develop intelligent systems which are increasingly relevant in education and training. For instance, Nuance, the high-tech company from Burlington (Massachusetts) [57], has implemented speech recognition software that can be used both by students and faculty. The application can transcribe up to 160 words per minute and is particularly useful for students who have limited mobility or struggle with writing. The available features also enhance word recognition and spelling. Teachers can apply the software to prepare homework and assemble and schedule recurrent tasks such as sending notifications or emails.

Another company, Knewton, is promoting its newest product, Alta [58], as a complete courseware solution that combines expertly designed adaptive learning technology with high quality openly available content. In other words, Alta helps identify drawbacks in a student’s knowledge by providing relevant coursework, and supports teaching activities at different educational levels.

Cognii is another provider of artificial intelligence by virtue of virtual assistants (VAs) that combine the powers of conversational pedagogy with the conversational AI technology [59]. Interestingly, the open-format applied to the VAs’ responses is claimed to improve critical-thinking skills. The VAs also provide real-time feedback, and individual tutoring, customized to the particular student’s requirements.

Querium, a successful start-up from Austin (Texas) [60], helps students master critical STEM skills by delivering a customizable STEM tutoring program of personalized and effective lessons which works on desktop computers and smartphones. Querium’s AI provides teachers with insights into the student’s learning habits and highlights areas in which the student should improve.

Century is another successful start-up, established in 2013 in London by Priya Lakhani [61]. Behind its success is a diverse team of teachers, technologists, neuroscientists and parents. Their platform utilizes data analytics and cognitive neuroscience to create personalized learning plans and reduce workloads for teachers. Moreover, the AI platform tracks student progress, identifies knowledge gaps and delivers personal learning recommendations and feedback. The teacher dashboard allows them to monitor individual student and whole-class performance.

These examples show only some of the possible applications of AI in the education sector. Nevertheless, we argue that the above allows us to claim that at a general level, the constellation of AI methods and tools leverage both learning and teaching. With time constraints, limited resources and the abundance of new, incoming knowledge, the engagement of artificial intelligence seems to be a must to deliver a competitive environment, facilitating both the learning and teaching processes.

According to the latest literature concerning the application of artificial intelligence in the education sector, computer-assisted tutoring represents the major applied field of AI [62]. At present, we can observe that the rate of development of intelligent educational software (IES) has increased due to the high and dynamic students demands [63]. 

The main problems are related to content flexibility and adaptability, and for reusability, sharing and collaborative development of the learning objects and structures [64]. Moreover, providing user-oriented content depends on three factors, namely the domain model, user model and instructional task model [65].
\subsection{Benefits}
The advantages of artificial intelligence applications in education are vast and varied. Here, everything can be considered to be beneficial if we are thinking of anything, for example a computer program, that can efficiently perform any task that would normally rely on the intelligence of a human. Based on the state-of-the art research in this area, we outline nine areas in which AI methods can bring added value for both learning and teaching activities [66].

The first benefit concerns automated grading which simulates the behavior of a teacher to assign grades to the answer sheets submitted by the students. It can assess their knowledge by processing and analyzing their answers, giving feedback and recommending personalized teaching plans.

Secondly, intermediate spaced repetition aims at knowledge revision when someone is just about to forget. It is worth noting that Polish inventor Peter Wozniak [67] introduced the SuperMemo application, which is based on the effect of spaced repetition. The app keeps track of what a user is learning, and when he/she is doing it. By applying AI techniques, the application can discover when a user is most likely about to forget something and recommend revising it.

Thirdly, feedback loops for teachers, aided by machine learning and natural language processing techniques, improves the quality of student evaluations. For example, a chatbot can collect opinions via a dialog interface similarly to a real interviewer but with a small amount of work required by the user. Moreover, each conversation can be adapted according to the student’s personality and provided answers. A chatbot can even formulate the reasons for particular opinions.

Fourthly, to support teachers in their classroom work, one can put into use virtual facilitators. For instance, at the Georgia Institute of Technology on Knowledge-Based Artificial Intelligence (KBAI) class, students were introduced to a new teacher’s assistant named Jill Watson (JW) [68], who has been operating on the online discussion forums of different offerings of the KBAI class since Spring 2016. JW autonomously responded to student introductions, answered routine, frequently asked questions, and posted announcements on a weekly basis.

In the fifth place, Watts introduced chat campus based on the IBM Watson cognitive computing technologies [69]. In brief, students at Deakin University have asked IBM Watson 1600 questions a week to learn the ins and outs of life on campus and studying in the cloud. Within 12 months of implementing Watson, due to the enhanced quality of the student know-how at Deakin, this ground-breaking solution has handled more than 55,000 questions from students. Furthermore, the school is progressing its use of Watson, broadening its capabilities and teaching the system to understand new sources of information.

Personalized learning is the sixth example of AI applications in the education sector. In general, it refers to a variety of educational programs in which the pace of learning and the instructional approach are customized and eventually optimized for the needs of each learner [70]. In particular, the content is tailored to the learning preferences and specific interests of each student.

The seventh example—one of the most promising—is adaptive learning (AL). While the traditional model of classroom education, continues to be very much one-size-fits-all, on the contrary, AI-powered AL systems are designed to optimize learning efficiency. For example, Yixue Squirrel AI (Yixue) collects and analyses students’ behavior data, updates learner profiles, then accordingly provides timely individualized feedback to each student [71].

Since cheating is a concern for all teachers, AI-powered anti-cheating systems have been presented as another (eighth) application of AI in the education sector. Proctoring is software which secures the authenticity of the test taker and prevent him/her from cheating as a proctor is always present during the test [72].

The last solution argued by Watts is data accumulation and personalization. For instance, learning grammatical rules can be aided by examples only from the domain being the subject of personal interest [73].
\subsection{Challenges}
There are several approaches to planning and organizing the implementation of AI methods in the education domain [74]–[81], but discussion about the essential challenges for decision-makers is still ongoing. To the best of our knowledge, the list of potential challenges that influence the implementation of intelligent technologies concern:
\begin{itemize}
\item \textbf{strategy} refers to a general plan of implementation to achieve one or more specific long-term goals accordingly to a schedule established and agreed with all interested stakeholders;
\item \textbf{organizational maturity} refers to its employees, processes and technology readiness and capability with respect to the adoption of artificial intelligence technologies;
\item \textbf{data governance}, refers to data principles, quality, meta-data, access requirements and data life cycle; since machines learn on the basis of data, data governance is a crucial facet of the implementation and further maintenance of AI;
\item \textbf{infrastructure}, being the combination of hardware and software systems, is particularly acute due to compatibility and integration issues.
\end{itemize}
As one would expect, it is important to establish a strategy that defines the goals with regard to the AI implementation and provides a means to manage them. The strategy itself might take the form of a mix of qualitative and quantitative approaches. The former aims to describe how the goals will be fulfilled, while the latter aims to decide if the goals are fulfilled and which goals are fulfilled. The fulfilment of the goals cab be expressed in quantitative numbers, or/and in qualitative terms.

In general, maturity is a synonym of “full development” or “perfected condition,” and since any organization is a living entity, it grows over a period of time and learn from its decisions and outcomes. Therefore, all organizations seem to be at some stage of maturity, striving forward to development and perfection. From a strategic point of view, we stress the importance of the high level of organizational maturity due to the changes spanning across core dimensions of strategic management such as: alignment, performance measurement and management, process improvement and sustainability. In the context of our study, maturity assessment should encompass external and internal benchmarks, describing the organization readiness and capability to adopt AI technologies.

Another challenge is data governance, which is related to the system of data organization, collection, control, storage, usage, archival and destruction. The path of setting up data governance is driven by a specific program, supported by particular policies and procedures, and communicated by organizational leadership and management. In general, the regulations must provide all of the necessary means to preserve the following generic requirements: accessibility, availability, completeness, accuracy, integrity, consistency, auditability, and security.

The last, but not least significant, challenge concerns the infrastructure which encompasses all of the hardware installations and the software entities. Recent advancements in the artificial intelligence technology landscape have introduced specific requirements toward hardware capacity and software capabilities. In an effort to integrate these cutting-edge technologies with the existing systems, one has to incorporate solutions that underpin a flexible and scalable end-to-end integration. Enabling on-the-fly software asset configurations and reconfigurations (in the case of enabling/disabling particular services) facilitates an “assembly-from-parts” model for implementing new and updating existing AI applications from a catalog of services.

We argue that the above-mentioned challenges are essential to take into consideration while preparing the discussed scenarios (see Section 4). Our research in essence attempts to identify and analyze the issues related to implementing AI in education, however it still lacks the empirical evidence which might at least confirm our perceptions of the studied phenomenon.

\section{Strategy implementation model}
Undoubtedly, artificial intelligence has had significant influence on various industries, leveraging their effectiveness, productivity and profitability. This also applies to the education sector which has been committed to several reforms, addressing key sectoral issues and including incorporating artificial intelligence tools and methods. 

It is believed that such a shift will bring a new perspective to many facets of existing learning and teaching techniques. Toumi, in his report from 2018, claims that [67]:
“(…) in the domain of educational policy, it is important for educators and policymakers to understand AI in the broader context of the future of learning. (…) As AI will be used to automate productive processes, we may need to reinvent current educational institutions. (…).”

This opinion and many others thereafter argue the need to better comprehend the impact of artificial intelligence on education. However, first we need to develop a model that will drive the process of implementation and result in the smooth deployment of AI systems. At this point, we argue that the below model is suitable under general contexts, and is thus applicable to any organization.

The five-stage process shown in Figure 3 organizes, arranges and systematizes the tasks into consequent groups. The particular tasks employed in each stage might cross other phases, to an extent which depends on the context of the implementation (e.g. strategy, organizational maturity, data governance). Therefore, the five stages are interdependent with each other, whereas the duration and labor intensity may significantly vary. 
\begin{figure}
\centering
\includegraphics[width=12.5cm]{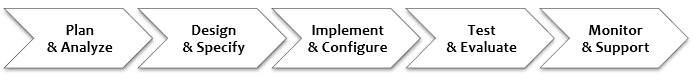}
\caption{General strategy implementation model.}
\label{fig:ProcessModel}
\end{figure}
The model of the implementation process consists of five stages:
\begin{enumerate}
\item \textbf{Plan and analyze} stage concerns all activities associated with the creation and maintenance of a plan which describes a list of steps with details of the timing and resources required to achieve the desired goals, along with the budget and general time frames.
\item \textbf{Design and specify} stage aims to prepare and establish the structure and organization of the system as well as to define the functional and non-functional requirements. In other words, a specification should address all enumerated goals in the first stage.
\item \textbf{Implement and configure} stage can be interpreted twofold; the system implementation is the process of creating its source code by the software developers, or the system implementation is installing and configuring the software applications.
\item \textbf{Test and evaluate} stage aims to ensure that the actual software is free of defects, and to check and determine whether it matches the expected requirements, as defined in the second phase.
\item \textbf{Monitor and support} stage concerns the surveillance process on measurable events regarding the performance of the system, as well as providing assistance and help to the users of the system required due to any issues and incidents encountered.
\end{enumerate}

Moreover, it should be emphasized that it is necessary to identify the specificities of AI software and evaluate existing resources, including hardware capacity, software compatibility and IT personnel. However, anticipating all of the relevant resources is only possible under exceptional circumstances exceptionally possible due to uncertainty regarding especially the human factor and data quality (used to teach AI algorithms).

\subsection{Implementation strategy configuration}
One should keep in mind that the strategy is a “how” the goals (objectives) will be achieved by the means (resources). In the context of this study, the strategy consists of the desired objectives and corresponding actions in order to implement an AI solution. Having said that, the structure of the implementation configuration is as follows:
\begin{itemize}
\item general purpose and vision,
\item scope, specific objectives with priorities and within time frames,
\item reference to the procedures used, regarding all five stages of the implementation model (see Fig. 3),
\item reference to the AI systems, services and tools which are existing and able to be included in the strategic plan, or in the case of developing a new system, a description of the possible software vendors, trusted advisors and consultancy agencies, and individuals experts, all with designated roles and responsibilities.
\end{itemize}

This perspective conceives the know-how as a highly deliberate, descriptive, and logical one, involving a sequential, rational and analytical collaborative work of all interested stakeholders. The configuration strategy is conceptually aimed at achieving the master plan by precisely following the implementation process. 

However, this form of strategy configuration can be mostly applied in organizations which at least manage projects in accordance with a policy that aims to identify and monitor the progress towards project performance objectives (level two).

\section{Exploratory Multiple Case Study}
Respecting the autonomy of higher education institutions, we would expect very individual approaches to deploying a single or suite of AI techniques for the university. The selected non-public universities that participated in the research are different in terms of geneses, formal statements as well as levels of computer infrastructure and areas of education. Before presenting the proposals for implementing artificial intelligence solutions in the inquired HEIs, short descriptions of the universities are delivered, focusing on their areas and levels of education. The state-of-the-art regarding the maturity of the implemented IT solutions are also discussed.

\subsection{Case settings}
\subsubsection{Case 1. WSZI in Wroclaw}
The university which is the subject of research of this article, namely The "Copernicus" University of Information Technology and Management, has been operating on the educational services market for nearly 20 years and is one of 384 non-public universities in Poland.

Since 2001, it has educated 4,000 graduates, mainly in the area of new technologies. Further observations and conclusions presented in the article will be based only on their own observations of business practice and the environment and competition due to the lack of data from external actors evaluating the quality, innovation and conditions for studying and living for students. Starting from 2019, the university has been conducting studies solely in the field of Computer Science in a full-time and part-time system at two levels, including post-graduate studies and courses.

The organizational structure of the university is a simplified structure with one department and full decision-making ability of the Chancellor, who is also the founder. His decisions are binding on financial, program and student matters. For the analysis of the organizational structure, one can state that WSIZ "Copernicus" omits the very important role of Rector. Usually, the rector in a non-public university is the person who deals with the organization of education and training, but his decisions are not binding without the authorization of the Chancellor. In addition, the described university is a unit educating a total of about 700 students in a full-time and part-time systems.

Among others: the department of the Rector's office, where there are too few employees and students. It is worth noting that if there were changes in the functioning of the organizational structure, it could be possible to reorganize the work and introduce intelligent technologies into these structures, introducing the innovation which would lead to the beginning of changes. 

The financing of a non-public university, which clearly affects the development and functioning of educational institutions, departs from the possibility of financing public entities. Obtaining funds or research grants is difficult due to the small number of students.

There is also no obligation to conduct research, which makes it more difficult to win in competitions. All retrofitting of the university halls are covered only from the profit earned, which consists of tuition payment and fees for program differences, conditions. When it comes to the issue of payment for tuition, the student has the privilege of financing science on the basis of a scholarships obtained, which are largely similar in terms of the procedures for obtaining funds in public universities.

A non-public university, even though it is a university, is also an enterprise. And just as a company will not exist without clients, a private university will not be able to break out of the educational market without the influence and opinion of its potential future students.

If we focus our conclusions on the differences in the functioning of these two types of universities, we should also mention the change in the law on higher education, which imposed student limits per lecturer on public universities. In WSIZ "Copernicus", the current number of students per lecturer is 25–30 students, which definitely departs from the provisions of law for public entities.

In addition, due to the lack of sufficiently qualified teaching staff and the Chancellor's desire to reduce employee costs, the university introduced the possibility of writing diploma theses to teachers with a master's degree, provided that formal care over diploma theses will be carried out by a promoter with a doctorate degree. However, he leads the student through the whole process of writing the work, advises him, gives his opinion on his progress and is listed in the diploma documentation. The content supervisor, who is subject to the promoter, is a person with a doctorate degree or above. Such a solution is allowed by the applicable law and applied in WSIZ "Copernicus", however, there is no information whether and to what extent public HEIs use it.

Looking at the above information about the structure, method of financing and processes taking place in the area of education and work organization of the described non-public university, it can be stated that although it is an entity educating students similarly to state universities, it is significantly different from them.

Obviously, it is not disputed that issues such as legal regulations, the shape and form of the graduate and student documents, requirements regarding ECTS points, and reporting or deadlines, are almost identical. However, they are so individual in their approach to the quality of education and the candidate that each of them should be considered separately. In the same way, when designing and implementing business analytics solutions, it should be taken into account that what works in a given non-public unit does not necessarily have to serve another organization, especially a state one.

\subsubsection{Case 2. WSB University in Gdansk} 
The WSB Universities is the largest group of university schools of business in Poland. All of them are found in top positions in rankings of higher education institutions. Founded in 1994 by TEB Akademia, WSB launched its first field of study: “finance and banking”. Presently, the WSB Universities are located in 10 cities across Poland: Gdańsk and Gdynia, Bydgoszcz and Toruń, Wrocław, Chorzów, Opole, Poznań, Szczecin, and Warszawa.

The organization of this particular University is typical for HEIs in Poland. The faculties and departments are oriented on educational sectors and there are many supporting divisions mentioned in the characteristics of the previous University, but particular units cooperate strongly, including by utilizing common platforms (webpages) for students and academic staff. 

The WSB educational offer covers both bachelor’s and master’s programs. The duration of the former is 3 years (6 semesters) while the latter is 2 years long. It is worth noting here that a student can pursue second-cycle programs in a field that is not directly related to that of their first-cycle degree. In that way, one can extend their area of professional expertise and hence enhance their employability. Each program is focused on developing practical competencies and soft skills.

Each university is governed by their own statutes and regulations, but are integral to the make-up of the ownership authorities. Moreover, each individual university has its own internal procedures, regulations, and local authorities. The university is governed by the rector along with the chancellor. At the head of each faculty are the dean and vice-deans, supported by the proxies.

University Faculties organize teaching and research into individual domains, or groups of subjects. Their work is normally organized into subdivisions called departments. The University's administrative and support departments support the running of the University and contribute to research, teaching and international cooperation with other universities worldwide.

The organization claims to maintain a high performance culture that is inspirational and motivating, providing internal and external funds to support faculty staff development. The aim of each subsidiary is to provide an attractive, sustainable, vibrant, and accessible campus, upheld by a contemporary virtual environment in which students, staff, and engaged stakeholders can interact, and share information and knowledge.

\subsubsection{Case 3. Jan Wyżykowski University in Polkowice.}

Jan Wyżykowski University was created thanks to the cooperation of Polkowice Commune and Polkowice County. The founder of the University is the ZAMPOL company, which Polkowice Commune and Polkowice County own all of the shares of. Currently, UJW runs studies within the Bachelor’s, Engineer’s, Master’s programmes and postgraduate studies and courses of Education offerings: Bachelor’s studies (3-year studies) in the fields of: Administration, Pedagogy and Management. Engineer’s studies (3.5-year studies) in the fields of: Information Technology, Mechatronics, Logistics, Mining and Geology, Production Management and Engineering. Master’s studies in the fields of Management (2-year), Mechatronics (1.5-year) and a uniform (5-year studies Law) are the newest in the offer. 

All educational majors are offered as extramural studies, so basically, students gain practical experience working for different companies. The second important feature of the university is that it acts in a very specific region known from its copper industry. Most of the students are employed in companies connected with this industry so their professional expectations are strictly tied with copper infrastructure or reflect more general positions typical for medium-sized cities (administration and management).

Nowadays, the information systems present in UJW support typical transaction processes: student enrollment, planning and evidence of classes and some functions offering by online education. It is quite a well-functioning university information systems but without functionality specific for AI methods.

\subsection{Implementation strategies settings}
\begin{figure}
\centering
\includegraphics[width=12.5cm,height=21cm]{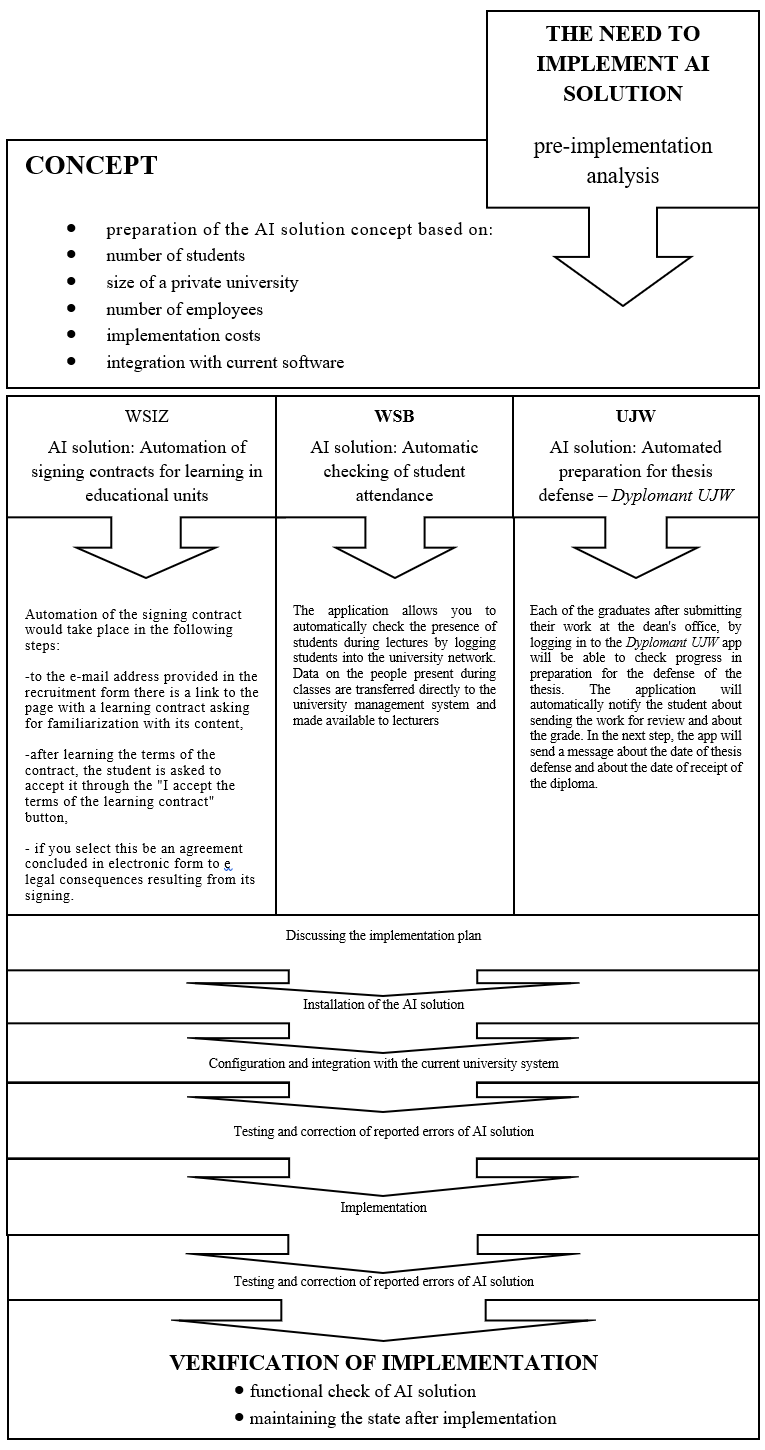}
\caption{Scenarios of implementation of selected AI methods for WSIZ, WSB and UJW.}
\label{fig:Scenarios}
\end{figure}

The solutions in Figure \ref{fig:Scenarios} use the latest technologies, which confirms that many repetitive tasks and procedures could benefit from the support of AI systems, thus offering new development opportunities and fields of study for higher education.

It is important to remember that there are many often more complex and non-standard procedures at universities. The processes of obtaining the intended effects from the implementation of the proposed solutions based on AI should be in accordance with applicable standards and the law regarding higher education and also take into account the capabilities of the university. Each replacement or support through the implementation of technological innovations is intended to shorten the time of document circulation, reduce administrative costs and improve the quality of education, which is certainly worthy of attention.

Naturally, introducing such a solution based on new technology solutions is not without its own inherent risks. Non-public universities would need to exercise extreme caution in protecting students’ personal data and would need some level of human oversight to monitor every AI method.

\subsubsection{Case 1. WSIZ university}
Building creative solutions into the work of a non-public university is also helpful in creating new products and services. The implementation of smart technology solutions at a private university will foster the building of scientific and educational progress, which will translate into satisfactory financial results and will strengthen the position on the non-public university market and introduce the element of innovation to the current activities of the unit. As part of the implementation work envisaged at the "Copernicus" University of Information Technology and Management in Wroclaw, it is planned to create a chatbot—an intelligent application for managing responses and relations with students and candidates for studies. 

This solution, in connection with the personnel problems of the non-public university described would answer simple questions about the dates of the sessions, exams, classes and inform about the recruitment schedule and the recruitment documents needed. As part of the planned implementation, it is also planned to reorganize the signing of learning contracts, which would consist in enabling their signing in electronic form via a link sent, generated by the internal university system, which is consistent with the provision in art. 60 cc.
Support for such a contract would take place in the following steps:
\begin{enumerate}
\item A link would be sent to the student’s e-mail address to the page about the learning contract with a request to read its contents.
\item After learning the terms of the contract, the student would be asked to accept it by clicking the “I accept the terms of the learning contract” button.
\item After choosing this option, the contract is concluded electronically.Afterwards, the student accepts the presented conditions, the confirmation along with the contract is sent automatically in a pdf file to the dean's employee. It is then printed and archived to the student's personal file.
\end{enumerate}

In addition, the plans of AI implementation in WSIZ also include the preparation and full implementation of a voice guide for students and candidates with impaired vision. The voice guide presents an educational offer and discusses payment, calls the phone number for the department, and checks the plan for that academic year.

The introduction of intelligent solutions for the work of the "Copernicus" University of Information Technology and Management in Wroclaw creates opportunities to facilitate procedures in the recruitment process, and reduce the use of office machines, which also affects the improvement of the environment. It relieves the work of the dean's office, which is the most important area in contact with students and candidates. 

Of course, apart from the positive aspects, one may ask the question whether intelligent technologies strive to improve human work and make it easier for them to complete certain processes to become more attractive on the market, or whether it is simply interference in human work and increasing unemployment. However, believing only in the good aspects of the case, we believe that the solutions and techniques of business intelligence will significantly improve the quality of the education process and assist in acquiring new students at the "Copernicus" University of Information Technology and Management in Wroclaw.

\subsubsection{Case 2. WSB Universities}
We have identified three administrative areas which are planned to be fully or partially automated by implementing artificial intelligence tools. The list is given below (not in order of priority).
\begin{enumerate}
\item \textbf{Grouping, sorting and responding to emails}. From our employees, we know that replying to emails is a time-consuming job. Moreover, repetitive email conversations are also frustrating and can be demotivating in the long-term. There are a number of tools which are being considered to be included in a pilot study, namely: AI Email Smart Answer [82], OMQ Reply [83], and Notion [84]. All of these tools are able to automate responses to emails and eventually replace the manual work of employees. Recurrent students’ requests are automatically recognized and answered by the system. Obviously, the list of features of the AI-powered solutions doesn’t end here. 
\item \textbf{Scheduling appointments}. Scheduling meetings with multiple students is a labor-intensive task. We believe that instead of manually responding via email to schedule an appointment and check the calendars of everyone involved, we could implement intelligent agents that detect and recognize certain phases in incoming emails, eventually proposing appointment times according to individual availability, and schedule appointments based on the attendees’ responses. Again, we have selected a few tools which are planned to be tested in a pilot study, namely: Julie Desk [85], ArtiBot [86], and Hendrix.ai [87]. Meetings are still a crucial method of organizing and planning work, but they are a waste of time unless one accurately captures what was discussed and agreed upon. By design, the agents require access to individual calendars, email accounts, social media profiles and location data to provide the necessary data to the AI-based inference engine. In return, they simply serve as virtual assistants, capable of preparing meetings, dialing into conference calls, turning on a video projectors, loading presentations, removing outdated and duplicate contacts, and many more otherwise tedious tasks.
\item \textbf{Customer service AI chatbot}. A chatbot is simply a software agent that can simulate a conversation with a user in natural language in a real time trough messaging applications, websites, mobile applications, or even over the telephone. The requests reported by the students dramatically increase before particular events, such as bachelor and master diploma exams, the beginning and the end of the academic year, as well as from the candidates who intend to become students. Therefore, customer service employees are regularly inundated with follow-up calls, support requests, frequently and repetitively asked questions, confirmation emails, complaints, and many more. To face these issues, IBM tells us that “chatbots can help businesses save on customer service costs by speeding up response times, freeing up agents for more challenging work, and answering up to 80\% of routine questions” [88]. 
\end{enumerate}

Moreover, unlike live agents (employees), software agents don’t need lunch hours or coffee breaks, and are not absent due to holidays, or illness, or any other natural disasters which can put human lives at risk. At the moment, our pilot study includes the following AI chatbot software to be tested: ActiveChat [89], Respond.io [80], and ChatBot [91]. We expect to uncover sizeable advantages by implementing chatbots across the organization. 

On the other hand, we are aware that it will not be an easy and effortless task. On the contrary, based on our learnings, we have identified and listed the following three major challenges, namely: security and privacy, obstacles and burdens due to polish language complexity, data input for machine learning algorithms. Undoubtedly, chatbots are one of the most promising enterprise AI technologies, however, achieving maximum business value from them requires from us extensive work and persevering determination.

\subsubsection{Case 3. UJW University}
Nowadays, the UJW educational unit is trying to be competitive as a local experimental HEI. Therefore, the educational offering relates to regional needs especially strongly connected with the copper industry and other companies located in the region. UJW was a leader of Erasmus grants devoted to improving the educational level through applying new teaching methods using AI (e.g. project DIMBI [92] and related [93]–[96]). Therefore all potential solutions in this HEI are strictly connected with the results of these projects extended by discovered niches in education. Three directions of AI techniques to embrace are:
\begin{enumerate}
\item \textbf{Using innovative methods of teaching selected courses}. At the beginning, practical abilities and skills refer to the data science and mechatronic majors.
\item \textbf{Smart agents} (SAs). SAs are used to automate administrative procedures with the aim to simplify daily tasks.  
\item \textbf{Chatbots}. Implementation of chatbots for communication with actual and potential students. This project is comparable to that presented in the previous universities. 
\end{enumerate}

Educating staff for the constantly changing trends and creating innovations both in the area of education and the functioning of a private educational unit translates into profit and brand visibility and the prestige of education. The search for creative solutions is the equation between the known, acceptable order and the chaos that innovation can bring. However, if innovation is treated as a course of action, as a way of managing, without worrying about disturbing relative stability, after a period of time, it can be seen that it has marked out new directions and possibilities of creating a competitive advantage. 

To sum up, we think that the automation of a universities’ administrative tasks and customization of their student-oriented activities are not only possible, but imminent. The goal of AI technologies is to make human-like judgments and perform tasks in order to downsize employees’ workloads. Today, academic communities are intensively developing and studying this field of AI applications. Indeed, the findings from research commissioned by the Microsoft show almost complete acceptance among educators that AI is important for their future— “99.4 percent said AI would be instrumental to their institution’s competitiveness within the next three years, with 15 per cent calling it a game-changer” [97]. 

However, in this paper, including the assumptions of the first pilot study, we “only” considered selected administrative areas, performed by the employees not directly responsible for education. Nevertheless, we believe that AI has far-reaching potential to change the way of teaching and learning. Undeniably, the incoming shift has its advocates and opponents whose proofs and claims should be always carefully judged.

\section{Conclusions}
As the technologies of artificial intelligence evolved, so did the domains and practices of their implementation in education. Current trends have imposed new requirements on the organization and management of both teaching and learning. There are three interrelated aspects of this, one of which arises from the recent advancements and innovations of the cutting-edge machine learning methods and Internet-of-Things devices. 

Secondly, the focus of both teachers and learners concern very large volumes of information and knowledge resources, freely available on the Web. Their growth in size and number seems to be an endless road to discover, but there is more and more evidence to help us pick the right direction. Thirdly, currently the competitiveness of higher educational institutions depends strongly on the increase in the effectiveness of learning methods, strongly supported by AI technologies and tools [98]–[100].

At a time when many education entities are stretched to capacity, and learners experience long wait times for on-site counseling, AI solutions could provide some facilities. It is therefore recommended that organizations should use solutions that are supported by the latest technological solutions, in order to improve the quality of education, and to minimize errors in the circulation of administrative documentation and the course of study. 

To sum up, we argue that the role and impact of artificial intelligence has increased in the education and learning contexts. The academic sphere is becoming more effective and personalized on the one hand, as well as global, context-intensive (multi-cultural) and asynchronous on the other. The intersection of three areas, namely data, computation and education has set far-reaching consequences, raising fundamental question about the nature of teaching: what is taught, when it is taught, and how it is taught.

\end{document}